\documentclass[9pt,twocolumnn,a4paper]{scrartcl}

\pdfoutput=1

\usepackage[hidelinks]{hyperref}

\usepackage{geometry}
\usepackage{microtype}
\usepackage{authblk}

\usepackage{graphicx}
\usepackage[abs]{overpic}
\usepackage{caption}
\usepackage{subcaption}

\usepackage{cuted}

\usepackage{amsmath, amsthm, amssymb}
\usepackage{amstext}

\usepackage{xcolor}

\usepackage{upgreek}
\newcommand{\fref}[1]{Fig.~\ref{#1}}

\newcommand{\be}{\begin{equation}}
\newcommand{\ee}{\end{equation}}

\newcommand{\ca}{c^\text{A}}
\newcommand{\cb}{c^\text{B}}

\newcommand{\fa}{f^\text{A}}
\newcommand{\fb}{f^\text{B}}
\newcommand{\fx}{f^X}
\newcommand{\caalpha}{c_\alpha^\text{A}}
\newcommand{\cbalpha}{c_\alpha^\text{B}}
\newcommand{\cabeta}{c_\beta^\text{A}}
\newcommand{\cbbeta}{c_\beta^\text{B}}
\newcommand{\caxi}{c_\xi^\text{A}}
\newcommand{\cbxi}{c_\xi^\text{B}}
\newcommand{\cachi}{c_\chi^\text{A}}
\newcommand{\cbchi}{c_\chi^\text{B}}
\newcommand{\cxchi}{c_\chi^X}
\newcommand{\pbab}{\bar{p}^{\text{AB}}}
\newcommand{\alphabxichi}{\bar{\alpha}_{\xi\leftrightarrow\chi}}
\newcommand{\onlinecite}[1]{\cite{#1}}

\newcommand{\pabn}{p^\text{AB}_{lmn}}
\newcommand{\pabnxichi}{p^{\text{AB}}_{lmn,\xi\chi}}
\newcommand{\pbanxichi}{p^{\text{BA}}_{lmn,\xi\chi}}
\newcommand{\alphan}{\alpha_{lmn}}
\newcommand{\alphatnxichi}{\tilde{\alpha}_{lmn,\xi\leftrightarrow\chi}}
\newcommand{\alphatnalphaalpha}{\tilde{\alpha}_{lmn,\alpha\leftrightarrow\alpha}}
\newcommand{\alphatnalphabeta}{\tilde{\alpha}_{lmn,\alpha\leftrightarrow\beta}}
\newcommand{\alphatnbetabeta}{\tilde{\alpha}_{lmn,\beta\leftrightarrow\beta}}
\newcommand{\rlmn}{\vec{r}_{lmn}}
\newcommand{\alternative}[2]{{#2}}
\newcommand{\kommentar}[1]{}

\begin{document}

%\title{\LARGE \hbox{Chemical short-range order in long-range ordered alloys} {\Large a simulation study using EAM potentials}}

\date{\today}

\title{Chemical ordering beyond the superstructure in long-range ordered systems}

%\author{Markus Stana$^1$, Manuel Ross$^1$ and Bogdan Sepiol$^1$}
%\ead{markus.stana@univie.ac.at}

\author{Markus Stana\protect\footnotemark,
Bogdan Sepiol}
\affil{Universit\"at Wien, Fakult\"at f\"ur Physik, Boltzmanngasse 5, 1090 Wien, Austria}

\author{Rafa\l~Kozubski}
\affil{M. Smoluchowski Institute of Physics, Jagiellonian University in Krakow, Reymonta 4 30-059 Krakow, Poland}

\author{Michael Leitner\protect\footnotemark}
\affil{TU M{\"u}nchen, Physik-Department E13, 85747 Garching, Germany}

%date{\today}

\twocolumn[

\begin{@twocolumnfalse}

\maketitle

\begin{abstract}
To describe chemical ordering in solid solutions systems Warren-Cowley short-range parameters are ordinarily used. 
However, they are not directly suited for application to long-range ordered systems, as they do not converge to zero for \alternative{distant coordination shells}{large separations}.
It is the aim of this paper to generalize the theory to long-range ordered systems and quantitatively discuss chemical short-range order beyond the superstructure arrangements. 
This is demonstrated on the example of a non-stoichiometric B2-ordered intermetallic alloy. 
Parameters of interatomic potentials are taken from an embedded atom method (EAM) calculations and the degree of order is simulated by the Monte Carlo method. Both on-lattice and off-lattice methods, where the latter allows individual atoms to deviate from their regular lattice sites, were used, and the resulting effects are discussed. 

\end{abstract}

\end{@twocolumnfalse}
]

\footnotetext{* markus.stana@univie.ac.at}
\footnotetext{$\dagger$ michael.leitner@frm2.tum.de}

\section{Introduction}

In elemental systems consisting of one type of atom, the structural configuration on the atomic scale in principle is defined by the crystal lattice. 
In contrast, in multi-component systems aspects of order give rise to additional degrees of freedom. On the one hand, this pertains to long-range order, where the crystal structure's sublattices have different elemental make-ups. On the other hand, short-range order corresponds to energetically preferred local arrangements within the stochastic occupations of the lattice sites by chemically distinct atoms. 

While the state of long-range order is determined by the sublattice compositions, the number of degrees of freedom of short-range order is in principle unbounded. 
In the simplest case of a binary system on a Bravais lattice, the Warren-Cawley short-range order parameters \cite{Cowley1950, Warren1951} quantify the probabilities of pairs of sites being occupied by a given combination of elements. These pair-level correlations are accessible directly in the intensity variations of diffuse scattering, and both elastic neutron scattering (for a review see Ref.\ \onlinecite{schweika1998disordered}) and X-ray scattering (see Ref.\ \onlinecite{schonfeld1999-2}) in the diffuse regime have been used for determining short-range order in a range of systems. 
 
Determination of long-range order via the intensities of Bragg and super-structure peaks is a routine experimental procedure \cite{xiao1995relationship, inden1990experimental}. In contrast, diffuse scattering experiments are much more labourious.
This pertains even more for studies of short-range order in long-range ordered compounds, where after the pioneering work by Georgopoulos and Cohen \cite{Georgopoulos1981} on $\beta'$-NiAl not much information has been published. The richness of aspects of short-range order that can generally be expected in ordered compounds, including both correlations of the occupations of sites within a given sublattice and on distinct sublattices, might be a reason for this. Note that even in the above-mentioned example of $\beta'$-NiAl the occupational disorder is restricted to a single sublattice, which allowed it to be described within the conventional Bravais lattice formalism. 

The outstanding properties of ordered intermetallics, including high yield strengths and high Young's moduli at low mass densities, corrosion resistance and high-temperature creep resistance \cite{sauthoff, pfeiler}, to name just a few, are due to the ordered arrangement of the atoms. However, lacking a detailed picture of local atomic arrangements, this has hitherto been considered only in terms of long-range order parameters. It is the aim of this work to take the first step towards a full model by providing the necessary formalism for describing short-range order correlations within long-range ordered systems, and to elucidate the accompanying phenomena by way of simulating temperature-dependent long- and short-range order. Specifically, we will consider the evolution of short-range order over an order-disorder transition in a non-stoichiometric B2 (CsCl) intermetallic alloy. We will assume a realistic potential in the vein of the embedded-atom method (EAM) \cite{Daw1984}. Such potentials are non-discrete, which allows us to compare simulations that explore the full classical parameter space including atomic displacements to simulations where the atoms are restricted to positions on an ideal lattice.

The paper will be organized in the following way: In Section \ref{sect:theory}, we will review the classical short-range order parameters and give a generalization for defining ordering on sublattices. We will detail the applied simulation technique in Section \ref{sect:simulation}, which will be followed by a discussion of the results in Section \ref{sect:results} and a brief summary in Section \ref{sect:conclusion}.

\section{Theory}\label{sect:theory}

Warren and Cowley \cite{Cowley1950, Warren1951, Cowley1960} introduced parameters to quantify short-range order in binary systems. They can be defined as
\begin{equation}
	\alphan = 1 - \frac{\pabn}{\ca\cb} \,,
	\label{eq:Wa-Co-class}
\end{equation}
where $\ca$ ($\cb$) is the overall concentration of atoms of type A (B) and $\pabn$ is the probability that a given pair of sites \alternative{that are in each other's $n$-th neighbouring shell}{separated by a lattice vector $\rlmn$} is occupied by distinct atoms.
%Neighbouring shells in this work are ordered in an ascending manner first by neighbour distance and then by largest vector component. \anmb{ich hab vor, niemals explizit so ein $n$ zu erwähnen, sondern immer konkret die miller-indizes, so dass der satz dann gestrichen werden kann} 
These parameters were originally introduced to describe SRO in systems where no LRO is present. Without symmetry breaking, short-range interactions lead to correlations that decay with distance, so that for large \alternative{indices of neighbouring shells $n$}{separations} the occupation of the sites is independent and therefore $\pabn=\ca\cb$ and $\alphan = 0$. 

Generally, the scattered intensity due to the sites of a Bravais lattice being occupied by different elements can be described by the expression\cite{schwartz2013diffraction}
\begin{equation}
	I(\vec{q}\,) = \sum_{\alternative{n,j_n}{lmn}} \alphan \cos( \vec{q}\cdot\vec{r}_{\alternative{n,j_n}{lmn}} ).
	\label{eq:isro}
\end{equation}
Here $\vec{q}$ denotes the position in reciprocal space, \alternative{$\vec{r}_{n,j_n}$ is the vector to the $j$-th position in the $n$-th neighbouring shell}{$\vec{r}_{lmn}$ are the lattice vectors}, and the intensity is understood in terms of the Laue unit \hbox{$\ca\cb(\fa-\fb)^2$}, where $\fx$ is the $\vec{q}$-dependent scattering factor. 
%\anmb{wie macht es schönfeld?} 
%\anmg{du meinst erst das $j$ un dann das $n$? ich find ja erst schalen und dann in den schalen durchnummerieren einleuchtendet, aber wenn du davon nicht abgehst ändere es halt} 
%\anmb{ich bin dafür, es so zu machen wie es schon andere vor uns gemacht haben, was auch immer da herauskommt. bei dir ist $j$ ein index zu $n$, was nicht dem tatsächlichen verhältnis entspricht. vielmehr sollte $n$ ein index zu $j$ sein, weil es das $j$-te atom in der $n$-ten schale ist. also am konsistentesten wäre für mich $\vec{r}^n_{j_n}$, wobei die summe über $n$ und $j_n$ geht. mit dieser notation wird auch klar, dass die beiden summationsindices verschiedene hierarchiestufen haben, die äußere summe muss ja über die schale gehen}
%\anmg{naja, schön ist anders, aber von mir aus}
%\anmb{na dann so wie schönfeld? also einfach 
%\be
%\sum_{\vec{r}}\alpha_{\vec{r}}\cos(\vec{q}\cdot\vec{r}),
%\ee $\vec{r}$ geht über alle gitterplätze und $\alpha_{\vec{r}}$ hat die symmetrie der punktgruppe? diese hierarchische unterteilung in schalen und positionen darin gefällt mir ja prinzipiell nicht, und ich glaub ich hab sie auch noch nie verwendet}

If the temperature compared to the ordering forces in the system is sufficiently low, a long-range superstructure lattice is formed. To give an example one may consider a B2 ordered system (ClCs structure), where the underlying body-centred cubic Bravais lattice dissociates into two simple cubic sublattices denoted as $\alpha$ and $\beta$. The long-range order parameter is given by the difference of the occupations of the sublattices 
\begin{equation}
\eta=(\caalpha-\cabeta)^2=(\cbalpha-\cbbeta)^2,
\label{eq:eta}
\end{equation}
where $\cxchi$ is the probability for a given site on sublattice $\chi$ to be occupied by an atom of element $X$.

With long-range order, the definition given in Eq.\ \eqref{eq:Wa-Co-class} will give non-zero values even for large \alternative{$n$}{$\vec{r}_{lmn}$}. For a stoichiometric system with perfect order and no defects the short-range order parameter $\alphan$ is either -1 or 1, depending on whether \alternative{$n$ indexes a neighbouring shell connecting}{$\vec{r}_{lmn}$ connects} sites on the same or on different sublattices. In this case Eq.\ \eqref{eq:isro} gives $\delta$-like superstructure peaks. The same is true for a long-range ordered system where the atomic concentration is non-stoichiometric, only that the excess atoms lead to the absolute value of the short-range order parameters being smaller than 1.

Specifically, assuming no short-range order we have
\begin{subequations}
\be
\pbab=\frac{\caalpha \cbbeta+\cbalpha \cabeta}{2}
\ee
for atoms within different sublattices and
\be
\pbab=\frac{\caalpha \cbalpha+\cabeta \cbbeta}{2}
\ee
\label{eq:prop-sub}
\end{subequations}
for atoms within the same sublattice. With the definition of $\eta$ and Eq.\ \eqref{eq:Wa-Co-class} we get
\be
\alphabxichi=\pm\frac{\eta}{4\ca\cb},
\label{eq:alphabar}
\ee 
where the positive sign is for same sublattices ($\xi=\chi$) and the negative sign for different sublattices. 

%In the special case of a triple defect B2 system, one sublattice is completely occupied by one type of atoms, while the second, hosting two types of atoms, can locally feature chemical short-range ordering. 

Under the presence of short-range order, the Warren-Cowley parameters for small \alternative{$n$}{$\vec{r}_{lmn}$} will reflect the actual short-range order, but for large \alternative{$n$}{separations} they will converge to the appropriate limiting value of the above two. The short-range ordering can now be quantified by the difference of the actual value to the uncorrelated expression of Eq.\ \eqref{eq:alphabar}
\be
\alphatnxichi=\frac{\caxi \cbchi-\pabnxichi}{\ca\cb}=\frac{\cbxi \cachi-\pbanxichi}{\ca\cb},
\label{eq:alpha-sublatt}
\ee
which indeed will converge towards zero after a small number of shells for non-critical temperatures. 
This allows the classical short-range order parameter to be written as a combination of a part that is due to long range order and a part that is strictly due to short-range ordering. 

Note that for the B2 case considered here, with two inequivalent sublattices, for given \alternative{$n$}{$lmn$} there are either two intra-sublattice short-range order coefficients or one inter-sublattice coefficient, corresponding to the cases
\begin{subequations}
\be
\alphan=\frac{\eta}{4\ca\cb}+\frac{\alphatnalphaalpha+\alphatnbetabeta}{2}
\ee
and
\be
\alphan=-\frac{\eta}{4\ca\cb}+\alphatnalphabeta,
\ee\label{eq:alpha-klass-sublatt}
\end{subequations}
respectively.
%\anmg{dass da das $p^n$ das $n$ oben hat und $\alphan$ das $n$ unten ist echt nicht schoen, aber schoenfeld schreibt übrigens $\alpha_{lmn}$}
%\anmb{ungeachtet meinen vorschlag oben: dann einfach alle spezifikationen von vektoren und schalen nach unten und alle atomsorten nach oben. also z.b. auch $\caalpha$, was dann auch die intuitive erklärung ``A auf $\alpha$'' hat}

\section{Simulation}\label{sect:simulation}

Configurations representative for the equilibrium configuration of the system were generated by the Metropolis-Hastings Monte Carlo method \cite{landaubinder2009}.
%The equilibrium atomic configuration of the system was simulated by means of a Monte Carlo algorithm numerically generating a Markov chain of configurations modified by stochastic exchanges of atomic positions in the lattice. In this way, time evolution of the density function of the configurations proceeds according to the Master equation and if the frequencies of the atomic-position exchanges fulfil the detailed balance condition, the density function converges to the Boltzmann distribution corresponding to thermodynamic equilibrium \cite{haider2007}. Practical realization of such simulations requires that interatomic interactions are determined by implementing particular potentials.
The internal energies of the specific configurations were computed according to the embedded atom method (EAM)-like potentials with long-range atomic interactions proposed by Ouyang et al.~\cite{Ouyang2012} for Fe--Al intermetallic alloys. For the two superstructures on the body-centred cubic lattice, the B2 phase with FeAl stoichiometry and the D0$_3$ phase with Fe$_3$Al stoichiometry, the resulting lattice constants, elastic constants, heat of fusion, point defect formation enthalpies, and the phonon dispersions were reported to agree very well with available experimental data.\cite{Ouyang2012,priv-Ouyang} In the following we will identify Fe with A and Al with B. %Unfortunately no short-range order parameters were derived in the original work. 

We implemented a grand-canonical Monte Carlo simulation, where a lattice site is chosen stochastically and its occupation is flipped with a probability that depends on the difference in the internal energies of the original and the flipped configuration and on a chemical potential. In order to achieve high efficiencies in the face of highly differing sublattice occupations, we allowed for a bias in attempt frequencies: specifically, if for a given sublattice the trial frequency for A$\rightarrow$B swaps is $\omega_{\text{A}\rightarrow\text{B}}$ and analogously for the reverse transition, then performing the exchanges with probabilities
\begin{equation}
p_{X\rightarrow Y}=\min\Bigl(1,\frac{\omega_{Y\rightarrow X}}{\omega_{X\rightarrow Y}}\exp\bigl(\frac{-\Delta H}{k_\text{B}T}\bigr)\Bigr)
\end{equation}
fulfills detailed balance.\cite{hastingsbiomet1970} Note that here $\Delta H$ incorporates both the differences in internal energy as well as the effect of the chemical potential. We implemented this bias by keeping lists of the sites on each sublattice occupied by either element. Choosing $\omega_{\text{A}\rightarrow\text{B}}/\omega_{\text{B}\rightarrow\text{A}}$ approximately as $\cbxi/\caxi$ (but fixed for the duration of the simulation) maximizes the acceptance rate and therefore the efficiency. We ensured the correct composition of the system via a temperature-dependent chemical potential. We simulate a physically meaningful evolution of time by incrementing at each trial with an exponentially-distributed random variable with a parameter according to the instantaneous trial frequency, which depends on the sublattice occupations.

We considered two statistical ensembles. The {\it on-lattice approach} constrains the atoms to their periodical lattice sites, so that the occupations of the sites are the only degrees of freedom. In the {\it off-lattice approach}, the atoms have the additional degree of freedom of deviating from their ideal positions. This was implemented by shifting random single atoms by random vectors and accepting the move with the Metropolis probabilities. As specific occupations of the lattice sites prefer specific off-lattice deviations (for instance, antistructure atoms of the larger element will lead to outward relaxations), allowing this degree of freedom slows down the occupational dynamics. We observed a slowing-down of about a factor of ten. Further, in the off-lattice simulations the lattice parameter was considered as an additional degree of freedom subject to Metropolis dynamics under zero external pressure, allowing the simulation cell to thermally fluctuate in size and to expand with increasing temperature.

We performed the averaging necessary to compute the simulated statistical parameters (such as the internal energy, lattice constant, sublattice occupations and pair distribution functions) by an exact temporal integral over the configurations as opposed to the conventional sampling approach. This was implemented by keeping records of the point in time a given quantity was changed the last time, and adding its value to the accumulating average with a weight given by the elapsed timespan when it is changed the next time. For the example of the pair distribution function, the neighbours of a flipped site are gone over, and the appropriate entry in the pair distributions is incremented proportional to the age of the pair, which is the minimum of the age of the two participating sites. Thanks to this approach and to the biasing discussed above, we can detect also correlations in defect species that become very small at low temperature, which would not be possible by a sampling approach. 

The size of the system with periodic boundary conditions for the simulation cell was chosen to be $32^3\times2 = 65536$ atoms. We have verified that this size was large enough to avoid serious finite-size effects, which is also corroborated by the sharpness of the ordering phase transition we will report below. A composition of $\ca=54\%$ was used for the calculations. In the on-lattice case, we used a lattice constant of $a=2.978$\,\AA{} at all temperatures, which is the low-temperature zero-pressure equilibrium value for this composition. No vacant sites were allowed.
%For each temperature the starting situation was that the Fe-sublattice was completely filled by Fe-atoms, while the excess Fe antisite atoms were randomly distributed on the Al-sublattice  
%\textcolor{red}{The average nearest neighbour distances for the off-lattice approach at this temperature was $1.0017\times d_\text{NN}$ for $A$-$B$ pairs and $0.9845\times d_\text{NN}$ for $A$-$A$ pairs, with $ d_\text{NN} = \sqrt{3}/2 a$ being the nearest neighbour distance in relation to the lattice constant $a$.} \anmb{\textbf{FIXME} was ist this temperature, außerdem überprüfen}

Ideally, in the off-lattice case atoms are kept on their position in the lattice only by the interactions with their neighbours. At elevated temperatures this would give a non-zero possibility for the system to transition towards a neighbouring local minimum of the high-dimensional potential landscape even under the absence of point defects, corresponding to diffusional jumps via some mechanism of direct exchange. This would break neighbourhood relations and prohibit the computation of order parameters. Therefore, during atom displacements we enforced the constraint that the position of each atom in relation to its eight direct neighbours has to conform to the order relations on the lattice, specifically, along each dimension the coordinate of a given central atom has to be larger than those of the four neighbours on one side and smaller that those of the neighbour on the other side. 

Of course, this constraint suppresses melting and leads to a different ensemble to be simulated. However, it is plausible that significant probabilities of such large displacements would lead to a melting of the unconstrained crystal, and indeed they appear only at temperatures where the lattice parameter has expanded by 3\% %and nearest-neighbour distances have aquired thermal broadenings of 10\%. 
and root-mean-square displacements with respect to the nearest neighbours have reached 18\% of the nearest-neighbour distance. 
Following the Lindemann criterion\cite{grimvall}, we assume that the range of existence of the crystal has been covered and stop our simulation at this point. In contrast, in the on-lattice ensemble the atoms are fixed to their lattice sites so that crystals of arbitrarily large temperature can be considered and simulated.

%If certain values remained constant or featured only fluctuations around a value after a certain number of MCs the system was considered to be in equilibrium. 

%For the on-lattice approach one Monte Carlo step takes less than one second on a standard PC. For the off-lattice approach with 70 relaxation attempts per atom, one Monte Carlo step takes 15 to 18\,h. It is, therefore, interesting to compare the results of both approaches, as in case that they are similar one can save a lot of computer time using much faster on-lattice approach.

\section{Results}\label{sect:results}
\begin{figure}[t]
	\hspace*{0.4cm}\includegraphics{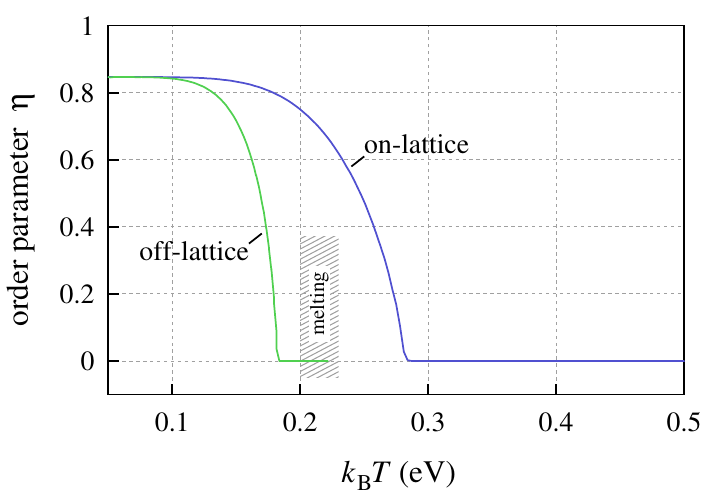}
	\caption{Long-range order parameter $\eta$ as a function of temperature for on- and off-lattice approaches. %\anmb{wieso hast du centering und die skalierung bei den abbildungen wieder eingebaut? erstens meinst du nicht 0.5textwidth, sondern columnwidth (ersteres nimmt anscheinend noch den abstand zwischen den spalten dazu, dadurch werden die abbildungen offensichtlich zu breit für die spalte), und zweitens hab ich die ohnehin gerade richtig breit gemacht, so dass man sie nicht skalieren muss und ein centering auch exakt keinen effekt hat}
}
	\label{fig:eta}
\end{figure}

As the most obvious aspect of chemical ordering in the system, long-range order was calculated as a function of temperatures in keeping with Eq.\ \eqref{eq:eta}. Results are shown in \fref{fig:eta}. They display the prototypical behaviour of Ising-like systems with only finite-energy excitations, so that the order parameter converges to its zero-temperature value faster than any power law. Specifically, below about $0.5\,T_\text{c}$ the system is practically fully long-range ordered, that is, the $\alpha$ sublattice is occupied exclusively by A atoms, while the surplus of A atoms leads to constitutional antisites on the $\beta$ sublattice.

Further, it is evident that the choice of the statistical ensemble has a significant effect on the bcc $\leftrightarrow$ B2 phase transition temperature, with $k_\text{B}T_\text{c}$ equal to $0.282\,$eV in the on-lattice case and about $0.183\,$eV in the off-lattice case. This is not surprising for two reasons: First, it is widely assumed \cite{Fultz2010247} that oscillatory normal modes in a given system get softer with disorder, thereby increasing vibrational entropy and lowering the free energy of the disordered phase. Of course, the system has this freedom only in the off-lattice approach, which would imply a lowering of the critical temperature compared to the on-lattice case. Second, the relaxations around antistructure atoms mentioned above decrease their formation energy compared to the on-lattice case. This results in disorder being also less costly in energy in case of the off-lattice approach.
%0.282 eV= 3273 K bzw. 0.194 eV = 2251 K
 
%\begin{align}		Das muss man nicht schreiben
%		&0.5\Bigg( \left(1+\frac{1}{2J}\right)\,\text{coth}\left( \left(1+\frac{1}{2J}\right)(x-x_0)\right) \notag \\
%		&- \frac{1}{2J}\, \text{coth}\left(\frac{1}{2J}(x-x_0)\right) \Bigg) - 0.5 \, ,
%\end{align}
%with $J = 0.07$ and $x_0 = 1$. 

\begin{figure}
	\hspace*{0.4cm}\includegraphics{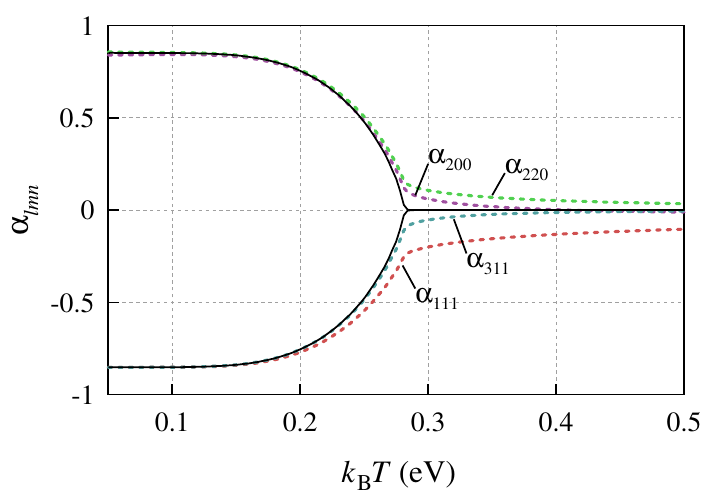}
	\caption{Classical short-range order parameter $\alphan$ as a function of temperature for the on-lattice approach. Values were calculated according to Eq.\ \eqref{eq:Wa-Co-class}. The solid black lines correspond to the values under absence of local short-range ordering according to Eq.\ \eqref{eq:alphabar}.
}
	\label{fig:alpha_1n2-bcc}
\end{figure}

\begin{figure*}
		\hspace*{0.4cm}
		\includegraphics{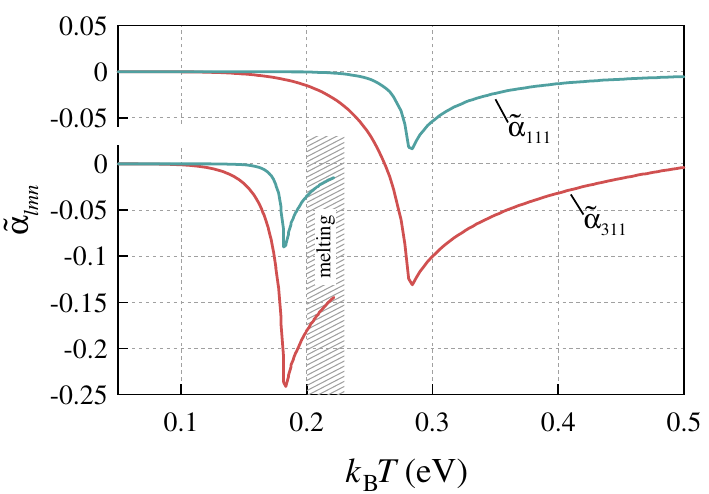} % [width=0.45\textwidth]
		\hspace*{\fill}
		\includegraphics{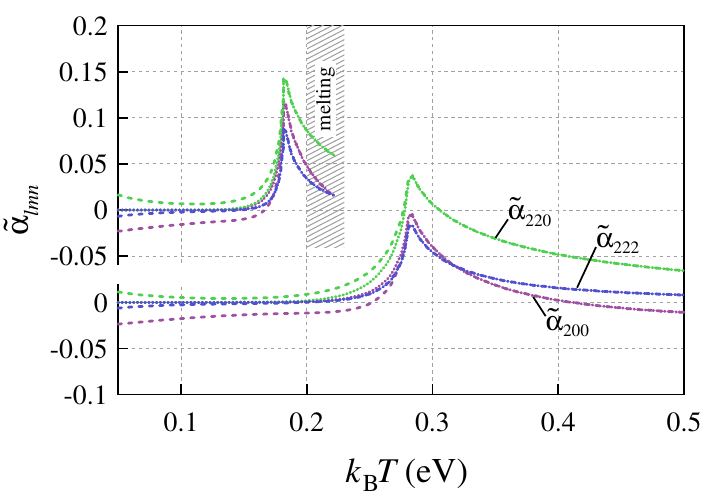}
		\hspace*{0.4cm}
	\caption{Sublattice-resolved chemical short-range order according to Eq.\ \eqref{eq:alpha-sublatt}. Inter-sublattice ($\xi \neq \chi$) parameters are illustrated in the left panel, intra-sublattice ($\xi = \chi$) parameters in the right, where dashed lines represent parameters for the majority atom sublattice and dotted lines parameters for the minority atom sublattice with structural antisites at low temperatures. Off-lattice values are shifted with respect to on-lattice values, and drawn only up to the melting of the system. 
%\anmb{streichen wir einfach die bezeichnung der vertikalen achse. das was ich da plotte ist eben $\alphatnxichi$, mit $\xi=\chi=\alpha$ bzw. $\xi=\chi=\beta$ unten und $\xi\ne\chi$ oben. und die sind ja unterschiedlich, schon allein weil sie für verschiedene $n$ definiert sind.} \anmg{Bitte schreib $\alphatnxichi$ statt $\alphan$ für die ordinate. Die achsenbeschriftung weg zu lassen würd ich ganz schlecht finden!!}
}
	\label{fig:subalphas}
\end{figure*}

The temperature-dependent short-range order parameters for the first four neighbouring shells were calculated according to Eq.\ \eqref{eq:Wa-Co-class}. The results are given in \fref{fig:alpha_1n2-bcc}. It can be seen that at low temperatures, where the system is well ordered, the classical short-range order parameters do not deviate appreciably from the appropriate non-correlated values according to Eq.\ \eqref{eq:alphabar}. This holds true even more for the inter-sublattice parameters, as the vanishing antisite concentration on the $\alpha$ sublattice $\cbalpha$ does not allow for any correlations to develop. For the intra-sublattice parameters very small deviations are discernible at low temperatures, which have to be due to short-range ordering of the constitutional antisites on the $\beta$ sublattice. In contrast, on the high-temperature side sizeable short-range correlations are stable to the highest simulated temperatures, which, as was to be expected, are largest for closest neighbours.

While the classical short-range order parameters are sufficient to calculate the total occupational scattering according to Eq.\ \eqref{eq:isro}, in the long-range ordered phase a large fraction of this intensity goes into the superstructure peaks. For discussing additional short-range order (or equivalently diffuse scattering), we also computed the sublattice-resolved short-range order coefficients according to Eq.\ \eqref{eq:alpha-sublatt}. Results are shown in \fref{fig:subalphas}. 

From the on-lattice results it is evident that above $T_\text{c}$ the two sublattices are equivalent, and therefore the pertaining short-range order parameters are equal, and, as can be deduced from Eqs.~\eqref{eq:alpha-klass-sublatt}, they are also equal to the classical parameters in \fref{fig:alpha_1n2-bcc}. Obviously also above $T_\text{c}$, the short-range order already hints at the eventual B2 symmetry breaking, as the inter-sublattice parameters are consistently negative, corresponding to a preference for pairs of unlike atoms if the sites are on different sublattices. Long-range order as illustrated in \fref{fig:eta} obviously is not sensitive to these issues. \textit{Mutatis mutandi}, the same holds for the intra-sublattice parameters. Here it is worth pointing out that, perhaps contrary to intuition, over a wide temperature range correlations over $\langle 220\rangle$ are stronger than those over $\langle 200\rangle$, even though the latter correspond to smaller separations.

The general behaviour of negative inter- and positive intra-sublattice short-range order parameters persists also some way below $T_\text{c}$, while the antisite concentration on the $\alpha$-sublattice is becoming increasingly smaller and correlations between these antisites vanish. The constitutional antisites on the $\beta$-sublattice, on the other hand, develop a characteristic signature corresponding to an increasing preference for nearest-neighbour pairs of unlike atoms within the sublattice. This suggests that the system would undergo D$0_3$ ordering at even lower temperatures. Further, it is remarkable that up to a non-linear rescaling of temperature, the off-lattice short-range order parameters behave quantitatively very similar to the on-lattice parameters. 

%\anmb{wollen wir jetzt wirklich noch eine abbildung wo eine reihe nahordnungsparameter für einzelne temperaturen, verschiedene modelle und verschiedene definitionen gezeigt werden?}
%\anmg{also den text bis hier, den du ergänzt hast finde ich einmal sehr gut. hab jetzt auch keine fehler gefunden. so wie du das mit den definitionen oben gemacht hast ist es ja schon evident, dass das das gleiche ist. ich bin mir jetzt auch nicht mehr ganz sicher und wiederspreche mir hier jetzt mit dem, was ich weiter oben angemerkt habe. wenn wir die abbilding, die ich drinnen hatte weg lassen, dann würd ich das zumindest im text noch stärker herausheben wollen, also unter der defintion 7a bzw 7b noch einmal dezitietr schreiben, dass der klassische $alpha$ parameter sich aus dem mittleren, der aus der lro entsteht und aus dem mit tilde zusammen gesetzt werden kann. wenn man das schön schreibt kann man auf die bilder denke ich verzichten. natürlich ist das aus den formeln ersichtlich, aber ich würde das trotzdem stärker durch den text noch einmal betonen}
%\anmb{feel free to do that. aber nicht einfach noch einmal wiederholen, weil das sieht ungelenk aus, sondern meinetwegen den text vor (7), wo ich das ja kurz auch im text erwähne, expliziter machen. und ach so, erst wo du es jetzt schreibst, sehe ich, dass dieses ``das gleiche'', von dem du hier schreibst, sich auf deine abb. 4a und 4b bezieht, wo aus 4a durch subtrahieren und umdefinieren von $n$ 4b wird. aber explizit erwähnt war das bei dir noch weniger.}

\begin{figure*}
	\hspace*{0.8cm}
  \includegraphics{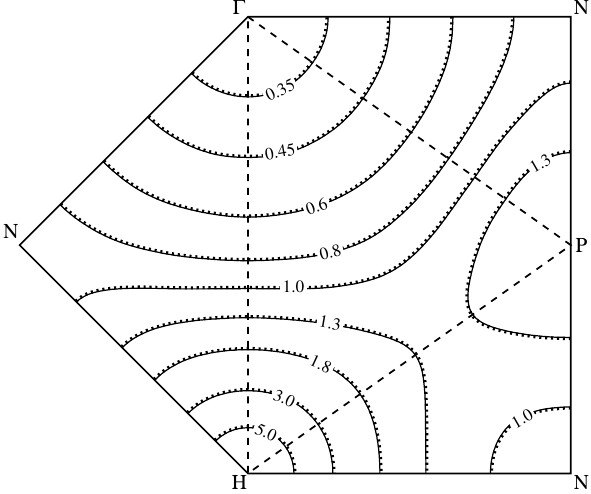}\hspace*{\fill}\includegraphics{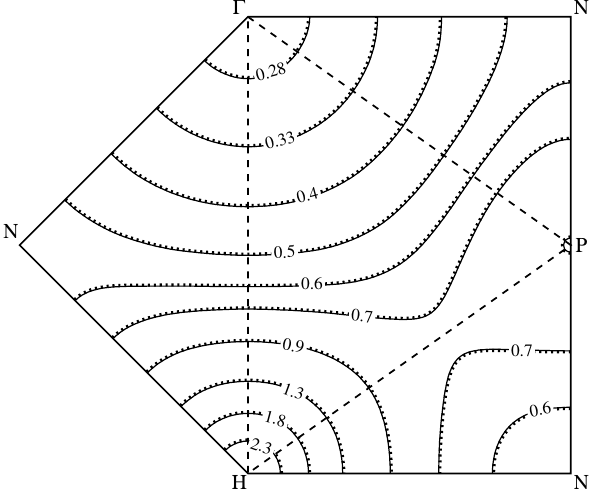}
	\hspace*{0.8cm}
  \caption{Diffuse scattering above (left) and below (right) the bcc $\leftrightarrow$ B2 transition, computed in the on-lattice approach for $k_\text{B}T=0.35$\,eV and $0.25$\,eV, respectively. The intensity is depicted over all facets of the irreducible element of the bcc Brillouin zone (a model of which can be prepared by cutting and folding back along the dashed lines). }\label{fig:diffint}
\end{figure*}

The short-range order diffuse scattering intensity as given by Eq.~\eqref{eq:isro} is depicted in \fref{fig:diffint}. For the two chosen temperatures both above and below $T_\text{c}$, the diffuse scattering looks qualitatively very similar. However, while the intensity stays roughly constant at the $\Gamma$ point, it decreases pronouncedly over the rest of the Brillouin zone during the ordering of the system. 
Note that the integrated scattering that is due to the sites of a lattice being occupied by idealized point-like scatterers (as assumed in the definition of short-range order intensity) does not depend on the arrangement of the scatterers. When long-range ordering is present, the $\delta$-like superstructure peaks compensate the reduction in short-range order diffuse scattering intensity. This is a consequence of a version of Parseval's theorem, which relates the squared $l_2$-norm of a function on a lattice (here the scattering lengths) to the squared $L_2$-norm of its Fourier transform (here the intensity).
%\anmg{kannst du da vielleicht normal-tauglich schreiben, warum das wichtig ist?} 
%\anmb{also das soll heißen, dass die intensität, die in die superstrukturreflexe geht, exakt die intensität ist, die in der diffusen streuung fehlt. und das folgt aus parseval. kannst du das normaltauglicher schreiben? ich könnte höchstens die nennung des theorems ohne seiner prosa-erklärung sehen lassen, aber hilft das etwas?} 
The Laue unit is chosen just so that the mean intensity is unity. If there is no long-range order, this directly holds also for the diffuse short-range order intensity. Under non-zero long-range order, on the other hand, the super-structure peaks account for a part of this scattered intensity, which in turn decreases the mean diffuse intensity. 
%\anmb{ich möchte jetzt ja keinen edit war anfangen, aber was genau stört dich am auskommentierten obigen absatz? ich find die erklärung besser und die argumente vollständiger als dein einzelner satz oben}\anmg{wo hab ich denn was von dir weg kommentiert?} \anmb{na die drei sätze direkt da drüber, angefangen mit ``The Laue unit\dots'', die ich jetzt wieder einkommentiert habe. jetzt ist es aber mit deinem ``When long-range ordering is present\dots'' gedoppelt, und außerdem hätte ich ``This is a consequence\dots'' ja auf die erhaltung der intensität bezogen, wovon die abnahme der diffusen intensität nur eine folgerung ist}

Apart from the decrease in intensity, the figures also demonstrate that the short-range order diffuse scattering in general has only the periodicity of the reciprocal lattice of the Bravais lattice. In the present case, only at very low temperatures, where because of the vanishing antisite concentration on the $\alpha$-sublattice all inter-sublattice parameters go to zero, the short-range order intensity would increasingly display the simple-cubic reciprocal-space symmetry of the B2 structure. 

%\clearpage

%\subsection{Vacancies}
%The vacancies are distributed on the sublattices according to the table:  ??? WELCHE?

%For the on-lattice approach the introduction of vacancies into the system increases the total energy of the system, as there is a smaller number of pairs diminishing pair energies. For the off-lattice approach on the other hand the introduction of vacancies leads to a lower total energy. This is in agreement with the observation of structural vacancies in Fe--Al and therefore an argument in favour of the off-lattice approach. It would be interesting to determine vacancy concentration in this way, especially at different temperatures, which would allow to determine the structural as well as thermal vacancies concentrations. Such a grand canonical, off-lattice approach will, however, be very expensive in terms of computation time. 

\section{Conclusion}\label{sect:conclusion}
%\anmb{also so wie die einleitung gehört das aus meiner sicht auch noch wesentlich umgebaut. gebrachte punkte: theorie, definition der $\tilde{\alpha}$, dh. trennung der echten nahordnung von der fernordnung. simulation eines EAM-modells, effiziente implementierung mit metropolis-hastings, demonstration des effekts von off-lattice auf übergangstemperatur. berechnung der nahordnungskoeffizienten, beispiel der theorie, berechnung der diffusen streuung}
%\anmg{ja da gebe ich dir recht. wenn der rest soweit fertig ist werde ich das einmal machen, dann kannst du es überarbeiten. oder willst du eine version schreiben und ich abreite drüber?}

%As stated in the introduction, short-range order parameters are essential for the correct interpretation of coherent time resolved scattering techniques, such as aXPCS. Only for few systems, like Cu-Au \cite{Leitner2009} the appropriate parameters can be found in literature (see \onlinecite{Schoenfeld1999}, we actually used the same sample for the aXPCS experiment as Sch{\"o}nfeld did for determining the Warren-Cowley parameters). For some very straightforward systems, a simple model can be found, as was the case for a Ni-Pt solid solution \cite{Stana2013}, where short-range order was clearly  due to repulsion of the minority atoms. In this work we could show that EAM potentials are a good way to find the desired short-range parameters for more complex systems by means of simulation, as was demonstrated on the the case of the B2 ordered structure Fe$_{0.54}$Al$_{0.46}$.

In the theoretical part of this work, we have shown how for long-range ordered systems the classical short-range order coefficients can be split into a term that depends only on the degree of long-range order and a term that is due to actual short-range order. The first term depends only on the sublattices the respective sites are on and does therefore not decay with distance. It is responsible for the sharp super-structure peaks. The second term represents deviations in the correlations in pair occupations from the long-range order term. For vectors in the structure's Bravais lattice, i.e.\ vectors that connect sites within a given sublattice, this short-range order term can further be written as a sum of parameters of correlations within the distinct sublattices.

Going beyond standard Metropolis simulations, we have presented an implementation of a Monte Carlo simulation in Hastings' framework, with a site-age based computation of the desired statistical quantities. These two complications allowed for an efficient study also of the very rare defects on the majority sublattice of an off-stoichiometric system. We simulated according to an EAM model of the Fe-Al system and demonstrated an increase in disordering temperature when constraining the atoms to the ideal lattice positions. 

Finally, as an exemplary use of the formalism introduced in the first part of this work, we computed the evolution of the distinct short-range order coefficients with temperature, showing an increase of short-range order correlations with decreasing temperature in the disordered phase, with a maximum at the ordering temperature, and a successive decrease as the correlations become long-ranged. We also illustrated the corresponding diffuse scattering. 

Additionally we could show that even tough on-lattice and off-lattice simulation approaches show qualitatively similar results for ordering parameters, the differences can not be neglected. 

Apart from exemplifying the concepts as done here, it is also conceivable to use simulations of diffuse scattering as substitutes for experimental data when short-range order information is necessary. Specifically, this pertains to the problem of data modeling in atomic-scale X-ray Photon Correlation Spectroscopy (aXPCS) \cite{Leitner2012, stana2014}, where in the simplest approximation\cite{Leitner2011} the measured $\vec{q}$-dependent correlation times are a product of the short-range order diffuse scattering and a factor depending on the jump geometry. Only in rare cases\cite{Leitner2009} experimental short-range order data is available for the same composition and temperature range\cite{Schoenfeld1999}, or simple qualitative models such as nearest-neighbour site exclusions\cite{Stana2013} can describe the data. In the general case, simulations of the diffuse intensity as presented here can serve as a starting point for interpretation\cite{stana2016}. Further, studies as presented here can also be used to test potentials against each other or against experimental data, as the diffuse short-range order intensity is sensitive to minute details of the atomic interactions.

%The considered EAM-potential leads to the formation of a super structure. This can be interpreted as attractive force between nearest-neighbour atoms of different type. Such a force, however, does not abruptly end after the first neighbouring shell. It was shown that an attractive force also exists for atoms of different type spaced by 100, while such atoms situated apart by 110 experience a repulsive force. This is represented by $\alpha_1^\beta < 0$ and $\alpha_2^\beta > 0$ for $T << T_\text{c}$. Such effects can not be represented by a nearest-neighbour interaction model for example.  

%It was therefore shown, that in the long-range ordered system under investigation, additional local chemical ordering (short-range order) is present. Neglecting SRO information in LRO systems is therefore generally not correct. 

%In a triple defect system at low temperatures, the inter-sublattice short-range order is completely dominated by the long-range order. Only  on  the  sublattice hosting the excess atoms short-range order is present between antistructure iron atoms. As soon as the phase transition takes place and long-range order disappears (decays), the intersublattice short-range order between nearest-neighour atoms becomes the dominant term.

\section*{Acknowledgements}

This research was funded by the Austrian Science Fund (FWF) contract P~28232. 
We also acknowledge the Scientific \& Technological Cooperation project number PL~12/2015 and funding by the Deutsche Forschungsgemeinschaft (DFG) through TRR 80.

For support with the code our thanks go to Stefan Puchegger. 

\bibliographystyle{mod_wmaainf} %gerplain, gerunsrt, geralpha, gerapali
\bibliography{sro}

\end{document}